\documentclass[prb,showpacs,superscriptaddress,twocolumn]{revtex4-2} 
\usepackage{graphicx,bm,amsmath,amssymb,psfrag}
\usepackage{color,xfrac,xcolor}
\usepackage{braket}

\begin{document}
	
	\title{Dynamic impurities in two-dimensional topological-insulator edge	states}
	\author{Simon Wozny}
	\email{simon.wozny@ftf.lth.se}
	\affiliation{NanoLund and Solid State Physics, Lund University, Box 118, 22100 Lund, Sweden}
	\author{Martin Leijnse}
	\affiliation{NanoLund and Solid State Physics, Lund University, Box 118, 22100 Lund, Sweden}
	\author{Sigurdur I.\ Erlingsson}
	\affiliation{School of Science and Engineering, Reykjavik University, Menntavegi 1, IS-101 Reykjavik, Iceland}
	
	\date{\today}
	
	\begin{abstract}
		Helical edge states of two-dimensional topological insulators show a gap in the density of states (DOS) and suppressed conductance in the presence of ordered magnetic impurities.  Here we will consider the dynamical effects on the DOS and transmission when the magnetic impurities are driven periodically. Using the Floquet formalism and Green's functions, the system properties are studied as a function of the driving frequency and the potential energy contribution of the impurities.  We see that increasing the potential part closes the DOS gap for all driving regimes.  The transmission gap is also closed, showing a pronounced asymmetry as a function of energy.  These features indicate that the dynamical transport properties could yield valuable information about the magnetic impurities.
	\end{abstract}

	\maketitle
	
	\section{Introduction}

Two-dimensional (2D) topological insulators (TI) host 1D counterpropagating (helical) edge states (quantum spin Hall effect) with linear Dirac-type dispersion \cite{xumoore,Bernevig2006, Konig2007, Qi2011}.
The edge states and the gapless nature of the linear spectrum are protected by time-reversal symmetry.
There is now a growing interest in utilizing the unique properties of 2D TIs and their edge states in applications such as, for example, spin filters \cite{Saffarzadeh2020}.
When considering applications and how to fabricate different devices \cite{breunig2021fabrication}, detailed understanding of the stationary and dynamic transport properties becomes increasingly important.

Backscattering of electrons in the helical edge states can occur in the presence of a spin-flip mechanism   \cite{wu+bernevig+zhang}.
Such mechanisms include, e.g., magnetic impurities \cite{Maciejko,altshuler13:086401,Tanaka2011,Kurilovich,vannucci2021}, interplay of nonmagnetic impurities and electron-electron interaction that leads to local magnetic moments \cite{novelli2019},  inelastic scattering due to electron-electron interaction \cite{schmidtglazman,Kainaris2014}, tunnel coupling to states in quantum dots \cite{Vaeyrynen2014}, coupling with nuclear spins \cite{Hsu2018}, and Rashba spin-orbit coupling \cite{strom,crepin,crepin2,Budich,Kimme2016}.
There are also efforts to investigate the effect of magnetic impurities in different geometries like, for example, TI rings \cite{Vezvaee2018}.
In some applications the goal is to open up a gap in the spectrum. This could be achieved by doping with magnetic impurities \cite{samarth2014,fritz2021}, but other mechanisms are possible, e.g.,\ currents that implicitly break time-reversal symmetry \cite{Balram2019}.
The influence of static aligned magnetic impurities on the density of states (DOS) has been investigated for 3D TIs \cite{Black-Schaffer2015} and we have previously considered the 2D TI static impurity case \cite{Wozny2018}.
Driven systems have also been studied, e.g.,\ a driven Rashba impurity coupled to helical edge states \cite{privitera21}, periodically driven via laser illumination \cite{huaman21},
and similarly a single driven nonmagnetic impurity \cite{Pradhan2019}.

In this study, we focus on the effects of aligned rotating magnetic impurities, which could be experimentally realized by ferromagnetic resonance \cite{Liu2006}.
Both the time averaged DOS and the transmission are calculated by using a Green's function (GF) approach.
The transmission is extracted via a scattering matrix formulation.
The behavior of the system is determined by the speed of driving relative to the time scale set by the inverse gap of the stationary system, and our method allows us to go from slow to fast driving.
We also investigate the effect of a  potential energy contribution from the impurities.
We explain the behavior in the DOS by comparing to the Floquet band structure of a homogeneous system.
The potential part leads to a flattening of both the DOS and the transmission.
While the effect of the impurities on the DOS is not visible for sufficiently large potential parts, the transmission does not approach the transmission of the clean system, since the potential part does not screen the magnetic moments that give rise to back-scattering.

The article is structured as follows. In Sec. \ref{sec:model} we introduce the system Hamiltonian and the GF formalism.
We also show how to extract the DOS and the transmission from the calculated GFs.
Section \ref{sec:results} presents and discusses the results, separated into three regimes of different driving speeds.
We close with a summary and conclusions in Sec. \ref{sec:summary}.

	\section{Model and methods}
	\label{sec:model}
	
\begin{figure}
	\begin{center}
		\includegraphics[angle=0,width=0.45\textwidth]{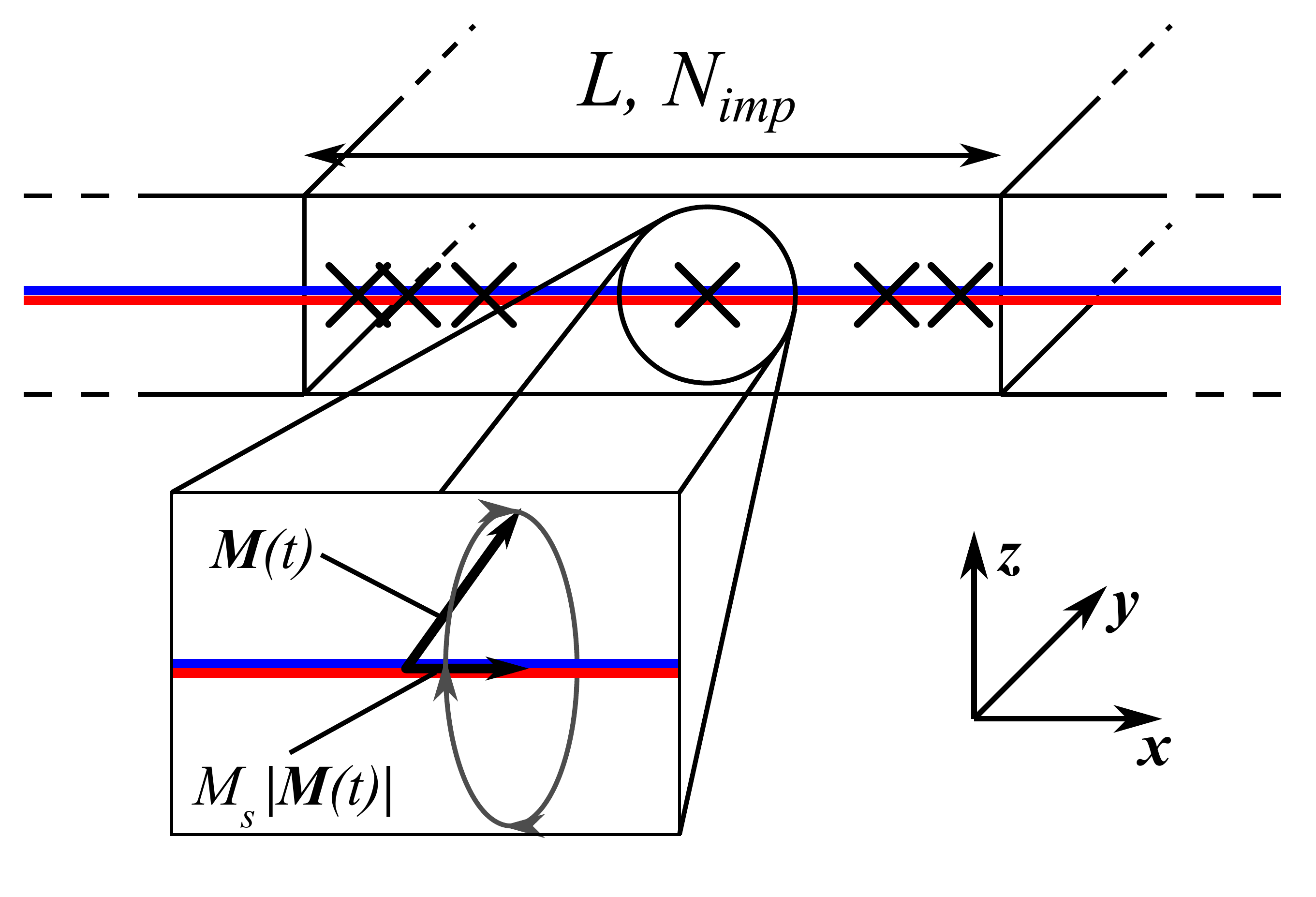}
		\caption{Schematic drawing of the impurity system of length $L$ with $N_{imp}$ impurities and a detailed view of a single impurity with indication of the dynamics of the magnetic moment. $\bm{M}(t)$ is the total magnetic moment of the impurity and $M_s$ is the ratio between the total magnetic moment and the static part pointing along the edge. \label{fig:intro}}
	\end{center}
\end{figure}

\subsection{Effective edge Hamiltonian}
The effective edge Hamiltonian in the Bernevig-Hughes-Zhang model \cite{Bernevig2006,Qi2011} with magnetically aligned rotating impurities precessing around the edge direction can be written as

\begin{eqnarray}
	H(t) &&= \hbar v_F k_x  \sigma_z+\sum_j 
	(V\sigma_0+\bm{M}(t) \cdot \bm{\sigma})\delta(x-x_j), \label{eq:edgeH}
\end{eqnarray}
where $v_F$ is the Fermi velocity, $V$  is the static potential part and the magnetic part of the delta impurities can be expressed as
\begin{eqnarray}	
	\frac{\bm{M}(t) \cdot \bm{\sigma}}{|\bm{M}|}=&&M_s  \sigma_y \nonumber \\
	&&+ \sqrt{1-M_s^2}  (\sin{(\Omega t)} \sigma_x + \cos{(\Omega t)} \sigma_z) \\
	=&&M_s  \sigma_y
	+\frac{\sqrt{1-M_s^2}}{2 i}(\sigma_x+i \sigma_z) e^{i \Omega t} \nonumber \\
    &&-\frac{\sqrt{1-M_s^2}}{2 i}(\sigma_x-i \sigma_z)e^{-i \Omega t}.
\end{eqnarray}
Here $|\bm{M}|$ denotes the strength of the magnetic impurities and $M_s$ is the ratio of the magnetic part pointing statically along the edge.
The sum runs over the impurity positions $x_j$.
We assume that all impurities have the same value of $V$ and $|\bm{M}|$ which are determined by an average over impurity positions relative to the edge state.
That means all impurities covered by the spatial extent of the edge states are taken into account with the same weight.
This procedure is performed analogously to the static impurity case and details can be found in the appendix of Ref.\ \cite{Wozny2018}.

In Fig. \ref{fig:intro} this type of system is shown schematically with an impurity region of length $L$ containing $N_{imp}$ impurities as will be used throughout the paper.

\subsection{Matrix Green's function}

Time independent transport can be described by a GF that only depends on one variable, i.e.\ the time difference $(t-t')$, and in the Fourier domain only one energy value is needed to specify the GF \cite{Doniach1998}.

For time-dependent systems the GF depends on two times variables, $t$ and $t'$. 
In the case of time-periodic driving the GF exhibits a particular periodicity that results in it being characterized by two energy parameters $E$ and $E'$, which are related by $E-E'=n \hbar \Omega$, where $n$ is an integer and $\Omega$ is the period of the driving \cite{Moskalets2012,Grifoni1998}.
A general harmonic driving term can be written as
\begin{equation}
U(t)=U_0+U_{+1} e^{i \Omega t}+U_{-1} e^{-i \Omega t},
\end{equation}
and the matrix equation for the Fourier components of the GF reads
\begin{equation}
	\tilde{G}_n(E)={} \tilde{g}(E)\delta_{n,0} + \sum_{m=0,\pm 1} \tilde{g}(E_n)U_m\tilde{G}_{n-m}(E), \label{eq:MatrixGF}  
\end{equation}
where we introduced the notation $E_n=E-n\hbar \Omega$.
The above equation can be written on explicit matrix form
\begin{widetext}
	\begin{equation}
		\begin{pmatrix}
			\ddots & \ddots & & & \\
			\ddots & 1-\tilde{g}(E_{-1})U_0 & -\tilde{g}(E_{-1})U_{-1} & &\\
			& -\tilde{g}(E)U_{+1} & 1-\tilde{g}(E)U_0 & -\tilde{g}(E)U_{-1} &\\
			& & -\tilde{g}(E_{1})U_{+1} & 1-\tilde{g}(E_{1})U_0 & \ddots\\
			& & & \ddots & \ddots
		\end{pmatrix}
		\begin{pmatrix}
			\vdots\\
			\tilde{G}_{-1}(E)\\
			\tilde{G}_{0}(E)\\
			\tilde{G}_{1}(E)\\
			\vdots
		\end{pmatrix}
		=
		\begin{pmatrix}
			\vdots\\
			0\\
			\tilde{g}(E)\\
			0\\
			\vdots
		\end{pmatrix}.
		\label{eq:matrixGF}
	\end{equation}
\end{widetext}
Given the free GF $\tilde{g}(E)$ the Fourier components $ \tilde{G}_{n}(E)$ of the GF are obtained by inverting the above matrix.
The free GF can be calculated explicitly and the static potential part can be absorbed into it.
Both have been calculated in Ref.\ \cite{Wozny2018}.
Note that the spatial dependence and spin has been suppressed and in principle each block in Eq.\ (\ref{eq:matrixGF}) has dimension $2N_\mathrm{imp} \times 2 N_{imp}$, reflecting the number of impurities and spin.

For the transport calculations we are interested in the GF connecting two arbitrary points $\bar{x}_1$ and $\bar{x}_N$ on the edge.
For that case the spatial dependence results in an additional summation and Eq. \eqref{eq:MatrixGF} reads 
\begin{eqnarray}
	\tilde{G}_n(E;\bar{x}_1,\bar{x}_N)&&=\tilde{g}(E;\bar{x}_1,\bar{x}_N)\delta_{n,0} \nonumber \\
	+\sum_{m=0,\pm 1}\sum_{k=1}^{N_{imp}}\, && \tilde{g}(E_n;\bar{x}_1,x_k)
	U_m \tilde{G}_{n-m}(E;x_k,\bar{x}_N)
	. \label{eq:endtoendGF}
\end{eqnarray}
The $\tilde{G}_n(E;x_k,\bar{x}_N)$ can be found by solving
\begin{eqnarray}
	\tilde{G}_n(E;x_i,\bar{x}_N)&&=\tilde{g}(E;x_i,\bar{x}_N)\delta_{n,0} \nonumber \\
	+\sum_{m=0,\pm 1}\sum_{k=1}^{N_{imp}}\, && \tilde{g}(E_n;x_i,x_k)
	U_m \tilde{G}_{n-m}(E;x_k,\bar{x}_N).
	\label{eq:imptoendGF}
\end{eqnarray}

The size of the matrices are determined by (i) the number of impurities and (ii) the number of Fourier components.  It is convenient to introduce the vector notation
\begin{eqnarray}
	\bm{b}_0&=&
	\begin{pmatrix}
		\tilde{g}(E;x_1,\bar{x}_N) \\
		\tilde{g}(E;x_2,\bar{x}_N)  \\
		\vdots  \\
		\tilde{g}(E;x_N,\bar{x}_N)
	\end{pmatrix}
\end{eqnarray}
and 
\begin{equation}
	\bf{v}_n =
	\begin{pmatrix}
		\tilde{G}_n(E;x_1,\bar{x}_N) \\
		\tilde{G}_n(E;x_2,\bar{x}_N)  \\
		\vdots  \\
		\tilde{G}_n(E;x_N,\bar{x}_N)
	\end{pmatrix}.
\end{equation}
Note, that all entries in the vectors are $2\times 2$ matrices reflecting the spin of the helical edge state.
The equation to solve is then 
\begin{equation}
	A
	\begin{pmatrix}
		\vdots \\
		\bm{v}_1 \\
		\bm{v}_0  \\
		\bm{v}_{-1} \\
		\vdots
	\end{pmatrix}
	=
	\begin{pmatrix}
		\vdots \\
		\bm{0} \\
		\bm{b}_0  \\
		\bm{0} \\
		\vdots
	\end{pmatrix} \label{eq:A}
\end{equation}
where $A$ is the full block-tri-diagonal matrix on the left hand side of Eq.\ \eqref{eq:matrixGF}.
The $N_{imp}$ sub-blocks in $\bm{b}_n$ and $\bm{v}_n$ are $2\times 2$ blocks, since they describe GFs for propagation from one specific impurity site to an arbitrary point on the edge, whereas the blocks of $A$ have dimension $2N_{imp}\times2N_{imp}$, since they contain propagations from every impurity site to every other impurity site.
Equation \eqref{eq:A} corresponds to the matrix equation in Eq.\ \eqref{eq:matrixGF} but with a slightly different right hand side and we can find the desired GFs by solving this matrix equation and inserting the results in Eq.\ \eqref{eq:endtoendGF}.
To solve Eq. \eqref{eq:A} we make use of the block-tri-diagonal structure and use the Thomas algorithm \cite{Quarteroni2010}.

	\subsection{Density of states}

	We want to find the DOS at the center of the impurity region to minimize tunneling contributions due to the finite size of the system.
	The tunneling length into the impurity region in the continuous limit is given by $l_\Delta = (\frac{N_{imp}}{L}\frac{M}{\hbar \nu_F})^{-1}$ and to minimize the tunneling contribution we choose $L = 8  l_\Delta$.
	The number of impurities that need to be placed in the numerical calculations is controlled by the ratio  $\frac{M}{\hbar \nu_F}$.
    The DOS can be calculated with the equal position GF, so we set $\bar{x}_1 = \bar{x}_N = L/2$ [the derivation following Eq. \eqref{eq:endtoendGF} holds for any value of $\bar{x}_1,\bar{x}_N$] and find the time averaged DOS via
	\begin{equation}
		DOS(E) = -  \mathrm{Im} \left( \frac{\mathrm{Tr} \,\left( G_0(E;L/2,L/2)\right)}{\pi} \right). \label{eq:dos}
	\end{equation} 
	
	\subsection{Scattering solution and transmission}
	Here we outline an extension of the derivation of Baranger and Stone in Ref.\ \cite{Baranger1989} for the scattering solution to the time-dependent case.
	The most general form of the scattering solution $\Psi_{\alpha,s}^{(+)}(x,t)$ in the presence of a driving term  $U$ is
	\begin{eqnarray}
		\Psi_{\alpha,s}^{(+)}(x,t)=&&\Phi_{\alpha,s}^{(+)}(x,t) \nonumber \\
	+\iint \mathrm{d}x' \mathrm{d}t' & &G(t,t';x,x') U(x',t') \Phi_{\alpha,s}^{(+)}(x',t'),
	\end{eqnarray}
	where $\Phi_{\alpha,s}^{(+)}(x,t)$ is the scattering solution in the {\it absence} of the driving and $\alpha$ denotes the incoming lead and $s$ is the spin \cite{Baranger1989}.
	Next, the equation of motion for the primed variables
	\begin{widetext}
	\begin{equation}
		-\mathrm{i} \hbar \partial_{t'} G(t,t';x,x') - (\mathrm{i} \hbar \partial_{x'} G(t,t';x,x') \sigma_z + G(t,t';x,x') U(x',t') ) = \delta (t-t') \delta (x-x'),
	\end{equation}
	\end{widetext}
	is used to replace the $G(t,t';x,x') U(x',t')$ in the integrand.
	Using the time-dependent Schr\"odinger equation for the clean system
	\begin{equation}
	    \mathrm{i} \hbar \partial_{t'}\Phi_{\alpha,s}^{(+)}(x',t')-v_F \sigma_z (-i\hbar \partial_{x'})\Phi_{\alpha,s}^{(+)}(x',t')=0,
	\end{equation}
	allows us to cancel all terms except the ones containing the partial derivative $\partial_{x'}\Phi_{\alpha,s}^{(+)}(x',t')$, which can be transformed into a boundary term by partial integration
	\begin{equation}
		\Psi_{\alpha,s}^{(+)}(x,t)=- \mathrm{i} \hbar \nu_F \left[ \int \mathrm{d}t'G(t,t';x,x') \sigma_z \Phi_{\alpha,s}^{(+)}(x',t') \right]_{x'=0}^{L}.
	\end{equation}
	Fourier transforming the above equation with respect to $t$ and using the periodic properties of the full GF results in
	\begin{equation}
		\Psi_{\alpha,s}^{(+)}(x,E)= \mathrm{i} \hbar \nu_F \sum_n
		\left[ G_n(E_n;x,x') \sigma_z
		\Phi_{\alpha,s}^{(+)}(x',E_n)\right]_{x'=0}^{L}, \label{eq:scatsol}
	\end{equation}
	where we used the Fourier decomposition of the GF and the convolution theorem; see Appendix\ \ref{app:FT_GF}.
	In this form we can project out the Floquet scattering matrix elements $S_{RL}^F(E_n,E)$ that describe the transmission of an incoming electron with energy $E_n$ from the (L)eft lead to the (R)ight lead at energy $E$, which results in
	\begin{equation}
		S^F_{RL}(E_n,E) = \mathrm{i} \hbar \nu_F \left[G_n(E;L,0)\right]_{\uparrow \uparrow}.
	\end{equation}
	The corresponding transmission function is given by
	\begin{eqnarray}
		T_{RL}^{(n)}(E)&=&|S^F_{RL}(E_n,E)|^2 \nonumber\\
		&=&\left(\hbar \nu_F\right)^2 \left|\left[G_n(E;L,0)\right]_{\uparrow \uparrow}\right|^2, \label{eq:trl}
	\end{eqnarray}
	similar to Ref. \cite{Moskalets2012} and in agreement with \cite{Camalet2003}, which also provides an expression for the current
	\begin{equation}
		\bar{I}=\frac{e}{h}\sum_n \int \mathrm{d}E \left[ T_{LR}^{(n)}(E) f_R(E) -T_{RL}^{(n)}(E) f_L(E)\right].
	\end{equation}
	The quantity $T_{RL}^{(n)}(E)$ describes the transmission of electrons from $\bar{x}_1=0$ to $\bar{x}_N=L$ under absorption/emission of $n$ energy quanta of $\hbar \Omega$. 
	In the impurity averaged case $T_{LR}^{(n)} = T_{RL}^{(n)}$ holds and we can simplify using a low temperature and small bias limit with bias $\Delta V$ around the chemical potential $\mu$ to find
	\begin{equation}
	 	\bar{I} = \frac{e^2}{h} \sum_n T_{RL}^{(n)}(\mu) \, \Delta V = \frac{e^2}{h} T (\mu) \, \Delta V,
	\end{equation}
	where we defined the effective transmission
\begin{equation}
		T(\mu)=\sum_n T_{RL}^{(n)}(\mu). \label{eq:T}
\end{equation}
Using Eq.\ \eqref{eq:trl} and impurity averaging the $T^{(n)}_{RL}$ we can calculate the transmission Eq.\ \eqref{eq:T} from the GFs obtained in Eq.\ \eqref{eq:endtoendGF}.
Note that $T_{LR}^{(n)} = T_{RL}^{(n)}$ only holds in the impurity averaged case, as an impurity distribution that is not symmetric around the center of the impurity region will break inversion symmetry.
This can be numerically verified once the Green's functions have been calculated.

Details and code implementation can be found in Ref. \cite{jupyternotebooks}.
	
\section{Results and discussion}
	\label{sec:results}

We present the results in three separate cases for different driving frequencies: fast driving $\hbar \Omega \gg \Delta $, intermediate driving $\hbar \Omega \sim \Delta $, and slow driving $\hbar \Omega \ll \Delta $, while keeping the magnetic impurity strength $|\vec{M}|$ constant.
Here $\Delta = \frac{N_{imp}}{L}\frac{|M|}{\hbar\nu_F}$ is the gap in the DOS expected for a homogeneous magnetic field in the $xy$ plane\cite{thesissimon}.
For each driving frequency we investigate the influence of a static potential strength $V$ on the transmission $T(E)$ and density of states $DOS(E)$.
In addition, we will consider the effects of a static magnetic component around which the dynamic magnetic part precesses.
The static part is chosen to be in $y$ direction, with the dynamic part rotating in the $xz$ plane and the contribution of the static part to the total magnetic moment given by $M_S |\vec{M}|$, as shown in Fig. \ref{fig:intro} and Eq. \eqref{eq:edgeH}.
This means that $M_s$ simply describes the proportion of the total magnetic moment statically pointing along the edge.
This choice is made to drive the magnetic moments between pointing in the $z$ direction, which does not induce a gap in the static case, and pointing somewhere in the $xy$ plane, which does induce a gap in the static case.
The choice is generic in the sense that a precession around any direction in the $xy$ plane with the same strength, frequency and static contribution will yield the same results.

The number of Fourier components for each driving frequency has been chosen by analyzing the Floquet quasi band structure of a clean system with a homogeneous, harmonically rotating field corresponding to the magnetization of the impurities.
The criterion here was to preserve approximate periodicity of the band structure in the observed energy range.
Details can be found in Appendix\ \ref{app:fbs} and the code used for this is available in Ref.\ \cite{jupyternotebooks}.

 \begin{figure}
 	\begin{center}
 		\includegraphics[angle=0,width=0.45\textwidth]{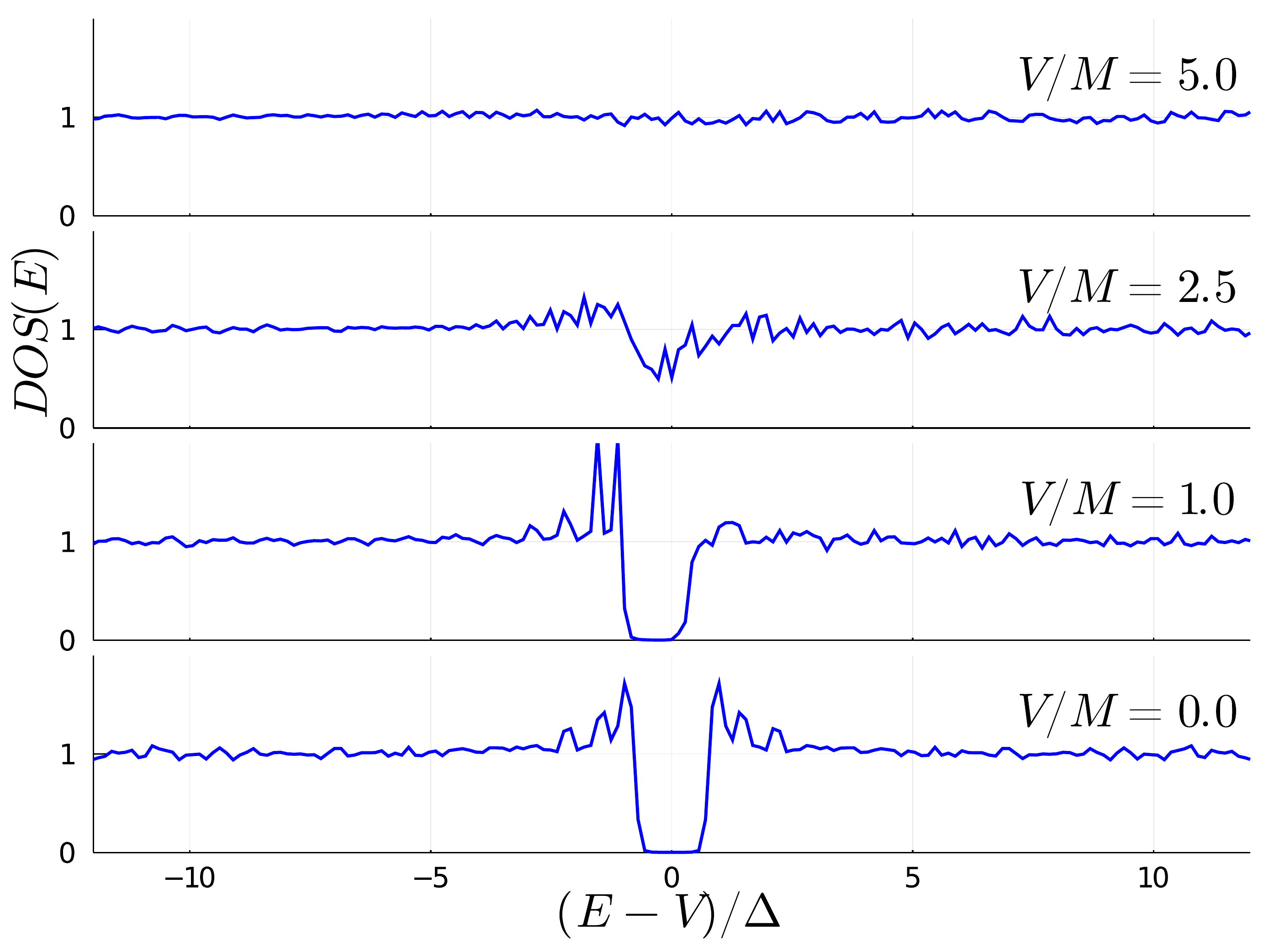}\\
 		\includegraphics[angle=0,width=0.45\textwidth]{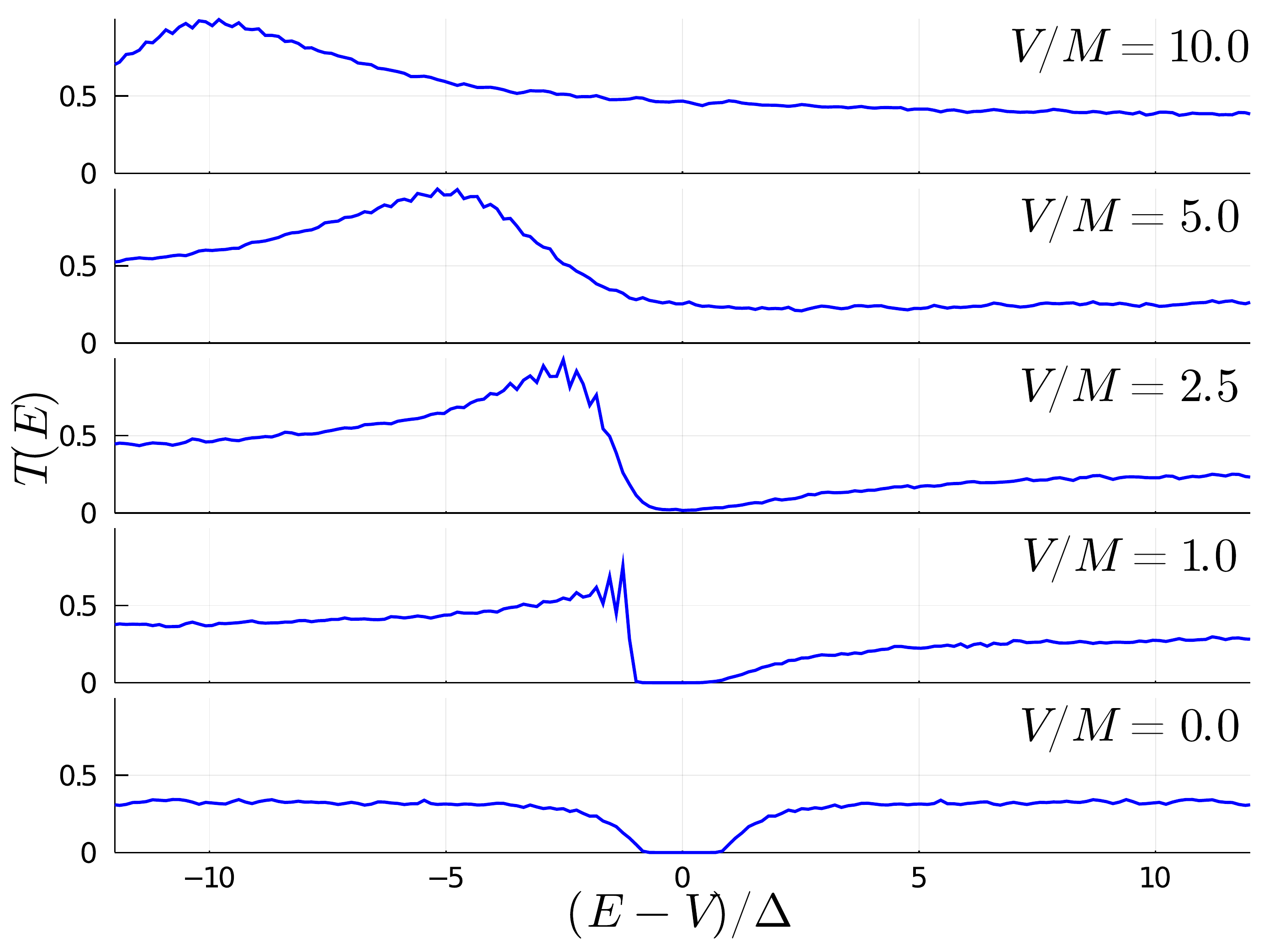}
 		\caption{$DOS$ and transmission $T$ as a function of energy for $M_s=1.0$ (fully static impurities) and different ratios of potential and magnetic part $V/M$, $|M|/\hbar\nu_F=0.2$, and $L/l_\Delta=8.0$ averaged over $N_{runs}=1000$ impurity configurations.}
 		\label{fig:static}
 	\end{center}
 \end{figure}

\subsection{Static case $\hbar \omega = 0$}

In Fig.\ \ref{fig:static} the results for $M_S=1$ are shown, corresponding to the purely static case.
The DOS shows a gap for purely magnetic impurities that is quickly filled as the potential part increases, in agreement with previous results \cite{Wozny2018}.
The transmission shows a similar gap that is filled with a rising potential part, but in contrast to the DOS the transmission is enhanced below the original gap position.
The asymmetry of the broadening of the peaks in the DOS can be understood by solving an infinite size system (with a fixed impurity density) in the Born approximation.
The self-energy in that case picks up a term proportional to $V \vec{M}/((1-V^2-M^2)^2+4V^2)$ that results in a bigger imaginary part for energies below the gap center as compared to above \cite{Wozny2018}.
In contrast to the DOS the transmission does not approach the clean system case for a rising potential part, i.e., there is no perfect transmission even for big potential parts implying that the magnetic part is not completely screened by the potential part and still enables backscattering in the system \cite{altshuler13:086401}. 

\begin{figure}
	\begin{center}
		\includegraphics[angle=0,width=0.45\textwidth]{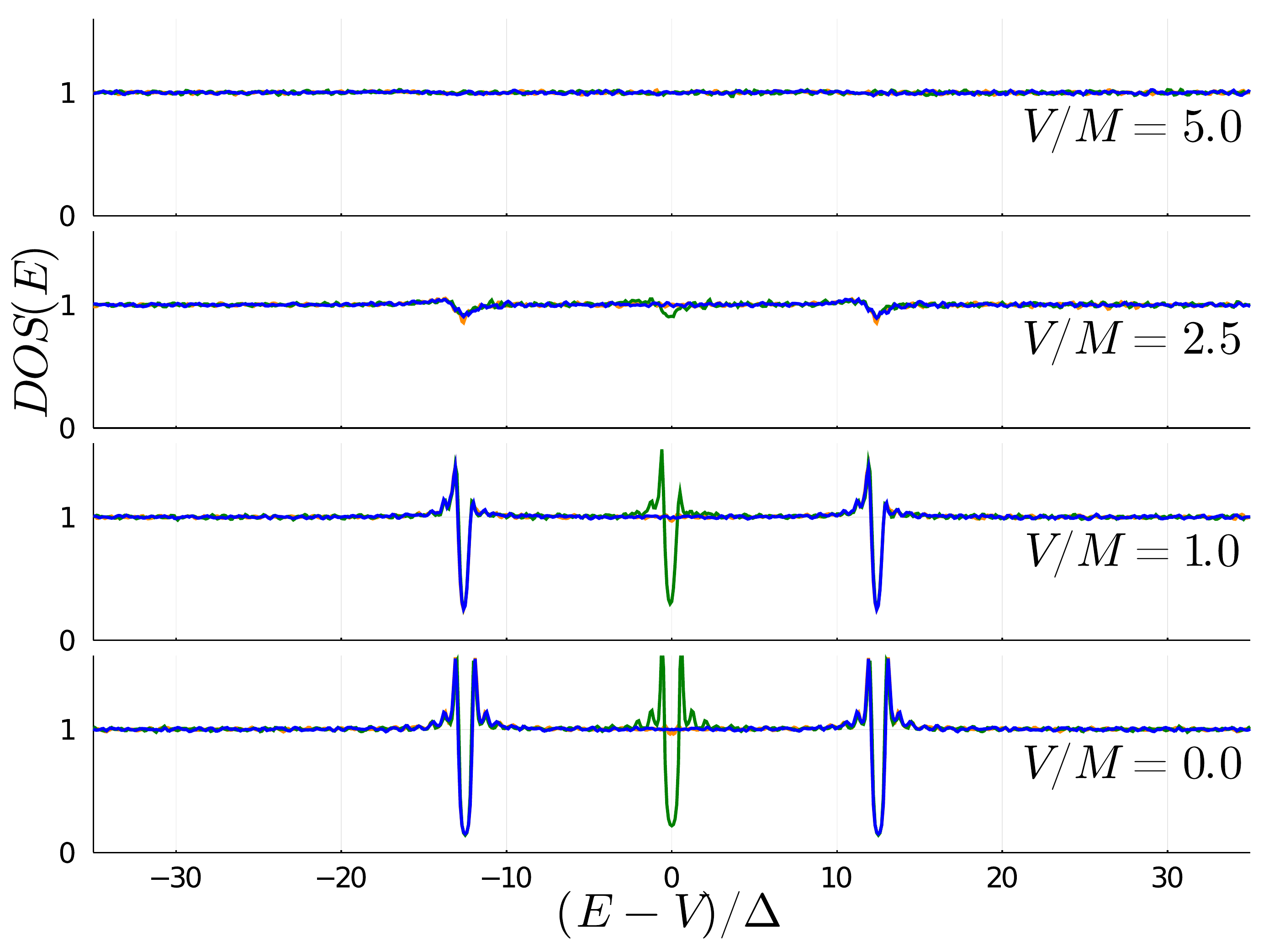}\\
		\includegraphics[angle=0,width=0.45\textwidth]{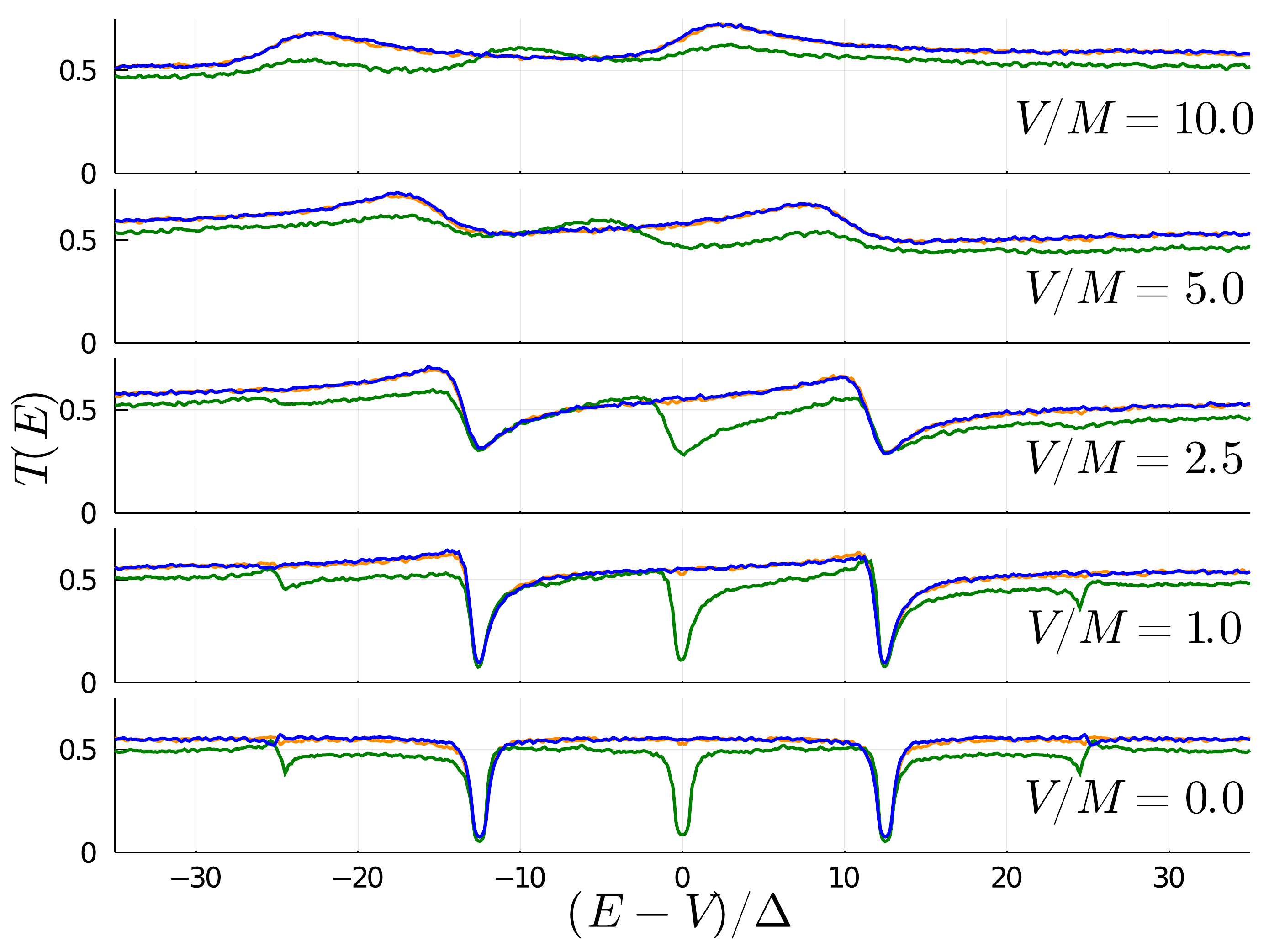}
		\caption{$DOS$ and transmission $T$ as a function of energy for different ratios of potential and magnetic part $V/M$ and $\hbar\Omega/\Delta=25$, $n_{max}=2$, $M_S=0.0$ (blue), $M_S=0.1$ (orange), $M_S=0.5$ (green), i.e., 10\%/50\% static part, $|M|/\hbar\nu_F=0.2$, and $L/l_\Delta=8.0$ averaged over $N_{runs}=1000$ impurity configurations.}
		\label{fig:hf}
	\end{center}
\end{figure}

\subsection{Fast driving $\hbar \omega \gg \Delta$}

Driving the impurities with high frequencies compared to the energy scale of the magnetic impurity strength results in distinct resonances in the DOS and dips in the transmission at energies corresponding to half the driving frequency as can be seen in Fig. \ref{fig:hf}.
These resonances can be understood if we consider a clean system with a homogeneous rotating field corresponding to the impurities. In that case we can calculate the Floquet band structure, which consists of copies of the original bands shifted by integer multiples of the driving frequency \cite{Grifoni1998}.
These sub-bands are coupled by the Fourier components of the magnetic field/impurities and thus the dynamic part leads to avoided crossings of the order of the magnetic impurity strength at first order subband crossings.
An example of the Floquet band structure along with the needed calculations can be found in Appendix \ref{app:fbs}.
Since in the fast driving case the size of the gap is small compared to the subband separation the features stay isolated.
As in the static case a rising potential part flattens the DOS quickly and fills the gap in the transmission while enhancing it just below the gap position.
This holds both for features due to the driving as well as for gaps induced by a finite static part.
The asymmetry here can be understood similarly to the static case where the off-diagonals of the center block lead to an asymmetric broadening depending on $V$, which also happens here for the different Floquet modes, leading to an asymmetry around each gap.
Comparing the blue traces in Fig.\ \ref{fig:hf} to the orange trace we can see that a small static part $M_S=0.1$ gives rise to a small feature in the center.
This feature is not as prominent as one would expect for the purely static case, potentially due to additional tunneling contributions, originating from the finite size of the system, that fill the small gap.
A static part $M_S=0.5$ results in a resonance of the same size as the dynamic features as can be seen in the green traces.
For the rising potential part the resonances behave similarly and are flattened in the DOS while the gap is filled and an area below the gap center shows enhanced transmission.
Also note that in this case the transmission exhibits features at energies $(E-V)/\Delta=25$ corresponding to the driving frequency.
They are related to second order crossings in the Floquet subband picture and are more visible and pronounced for bigger static contributions.
The static magnetic part does not directly affect the higher Floquet subband crossings, but due to the off-diagonal driving a higher effective magnetic part is picked up and thus higher order features are more pronounced.

\begin{figure}
\begin{center}
	\includegraphics[angle=0,width=0.45\textwidth]{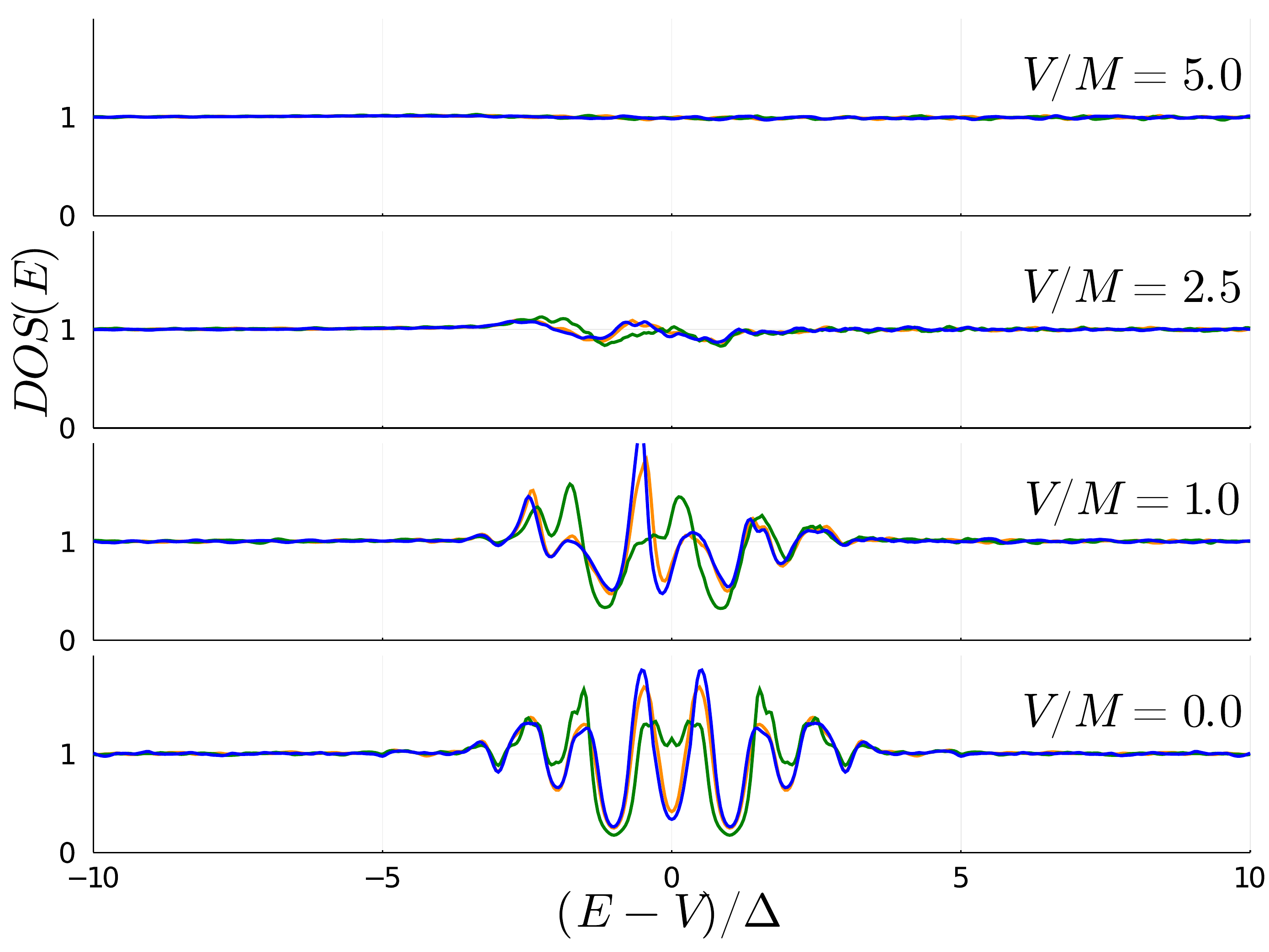}\\
	\includegraphics[angle=0,width=0.45\textwidth]{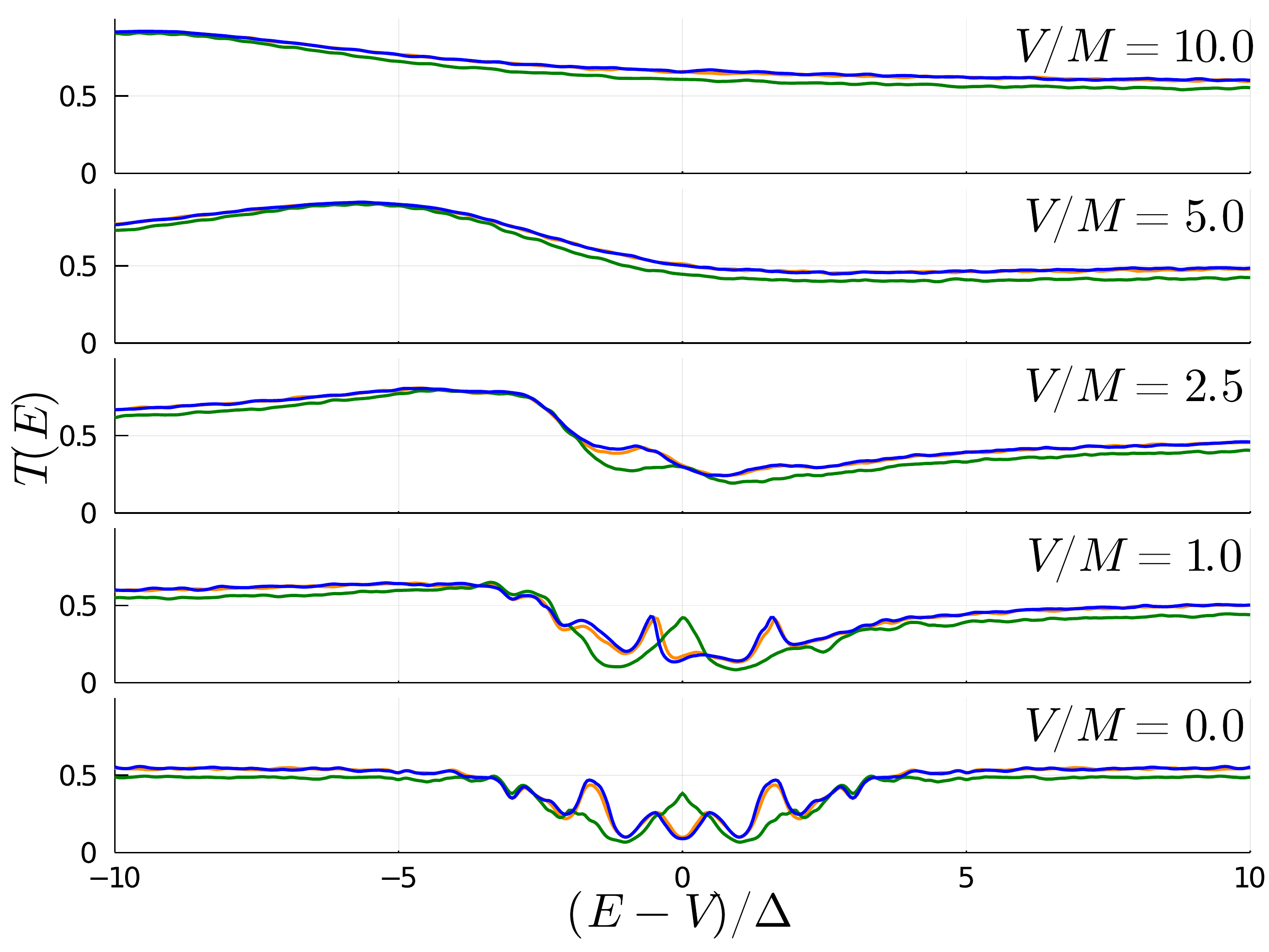}
	\caption{$DOS$ and transmission $T$ as a function of energy for different ratios of potential and magnetic part $V/M$ and $\hbar\Omega/\Delta=2$, $n_{max}=6$, $M_S=0.0$ (blue), $M_S=0.1$ (orange), $M_S=0.5$ (green), i.e., 10\%/50\% static part, $|M|/\hbar\nu_F=0.2$, and $L/l_\Delta=8.0$ averaged over $N_{runs}=1000$ impurity configurations.}
	\label{fig:res}
\end{center}
\end{figure}

\subsection{Intermediate driving $\hbar \omega \sim \Delta$}

When the driving frequency is of the same order of magnitude as the impurity strength the Floquet subband crossings are split further than the subband separation and the structures become nontrivial.
In Fig.\ \ref{fig:res} the DOS shows oscillations that are asymmetric around $E-V=0$ and become suppressed by a rising potential part.
A small static contribution leads to almost no difference (orange trace), while a larger static part leads to different behavior around zero and a different influence of the potential part (green trace).
Nevertheless, the potential part flattens the DOS quickly.
The transmission in the intermediate driving case behaves similar to the DOS; the driving leads to a small scale structure in the transmission around zero.
Again, a static contribution to the magnetic moment changes that structure significantly; in particular, it enhances the transmission around zero.
A rising potential part destroys the structure and flattens the transmission with an enhancement below the original structure similar to the high frequency case.

\begin{figure}
\begin{center}
	\includegraphics[angle=0,width=0.45\textwidth]{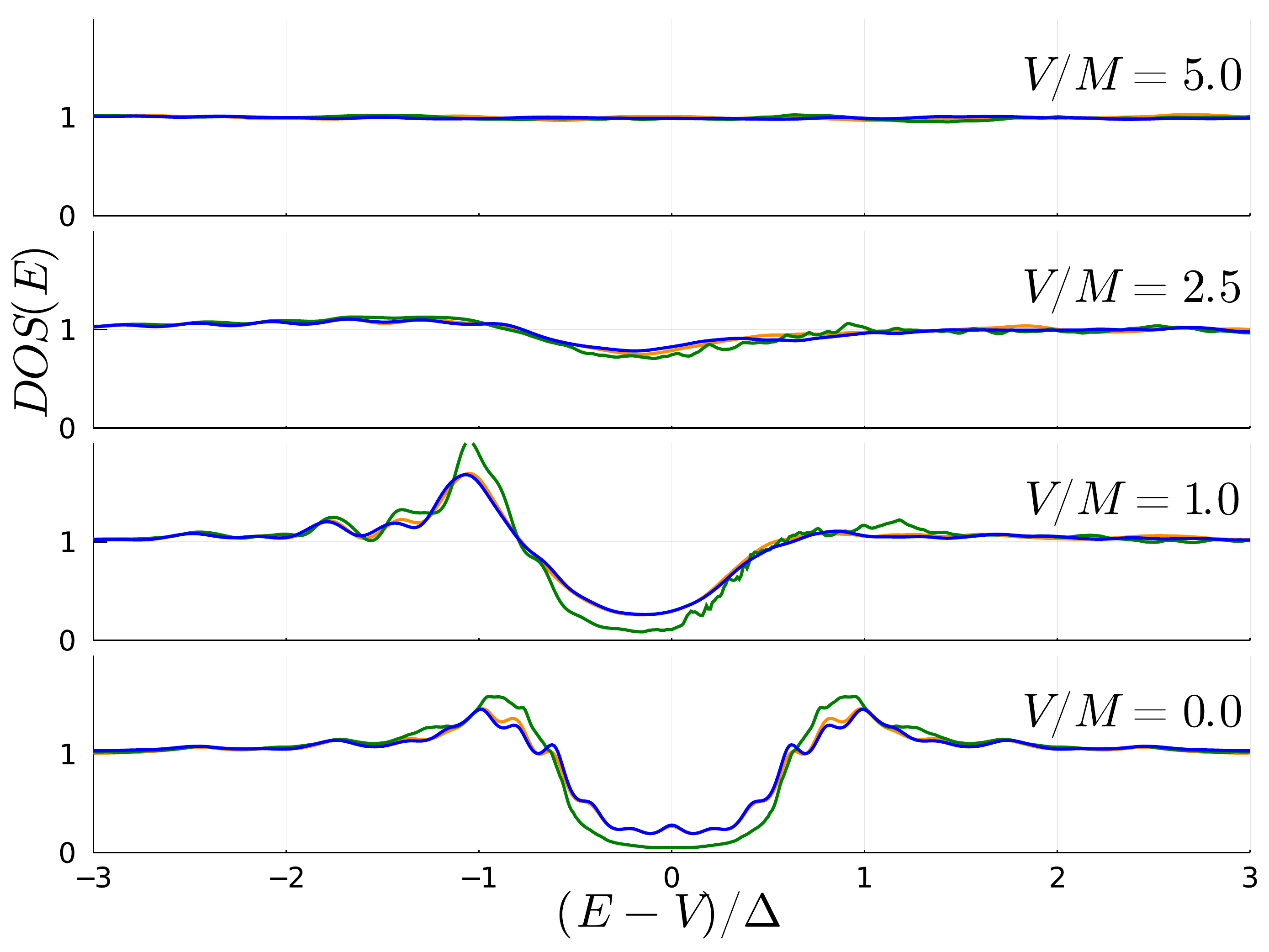}\\
	\includegraphics[angle=0,width=0.45\textwidth]{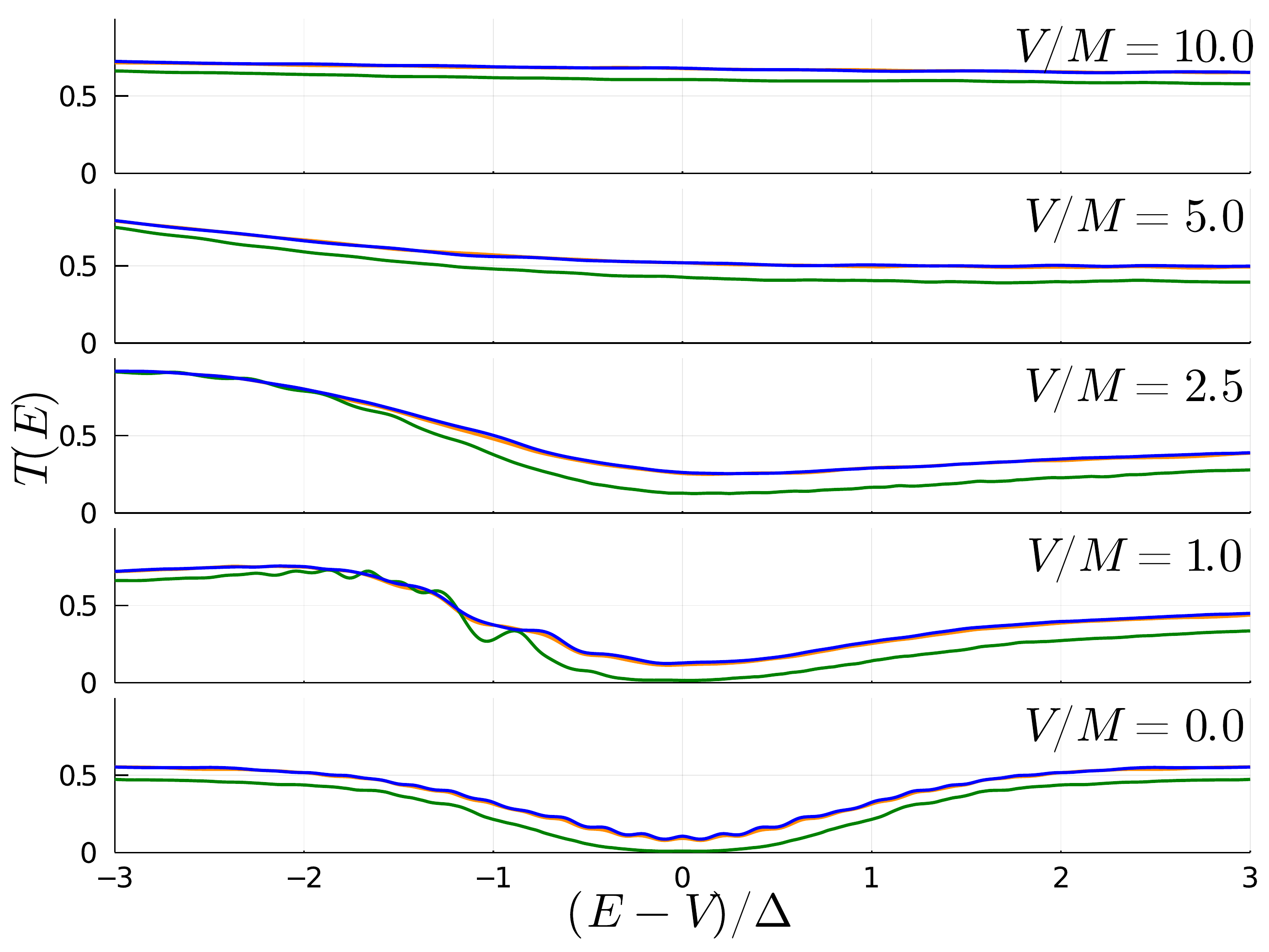}
	\caption{$DOS$ and transmission $T$ as a function of energy for different ratios of potential and magnetic part $V/M$ and $\hbar\Omega/M=0.2$, $n_{max}=20$, $M_S=0.0$ (blue), $M_S=0.1$ (orange), $M_S=0.5$ (green), i.e., 10\%/50\% static part, $|M|/\hbar\nu_F=0.2$, and $L/l_\Delta=8.0$ averaged over $N_{runs}=1000$ impurity configurations.}
	\label{fig:lf}
\end{center}
\end{figure}

\subsection{Slow driving $\hbar \omega \ll \Delta$}

For slow driving frequencies the results shown in Fig.\ \ref{fig:lf} exhibit similar behavior as in the other cases with regard to the potential strength. Increasing $V$ leads to a flat DOS and the features in the transmission are asymmetrically flattened. The shape of the transmission and DOS can be explained by assuming adiabatic driving, i.e., the response time of the system is much faster than the driving.
Then the average over one driving period corresponds to the average over results for static impurities with different impurity directions according to the rotation.
Hence for a bigger static contribution the wider gap for a small potential part as seen in the green trace is expected as well as the similar position of the outer edges.
With a rising potential part the gap is filled and the difference in the DOS becomes negligible.
Notably, already for $M_s=0.5$ the transmission in the gap center approaches zero, suggesting that the gap due to the static contribution is too big to be filled by the broadening due to the driving.

\begin{figure}
\begin{center}
	\includegraphics[angle=0,width=0.45\textwidth]{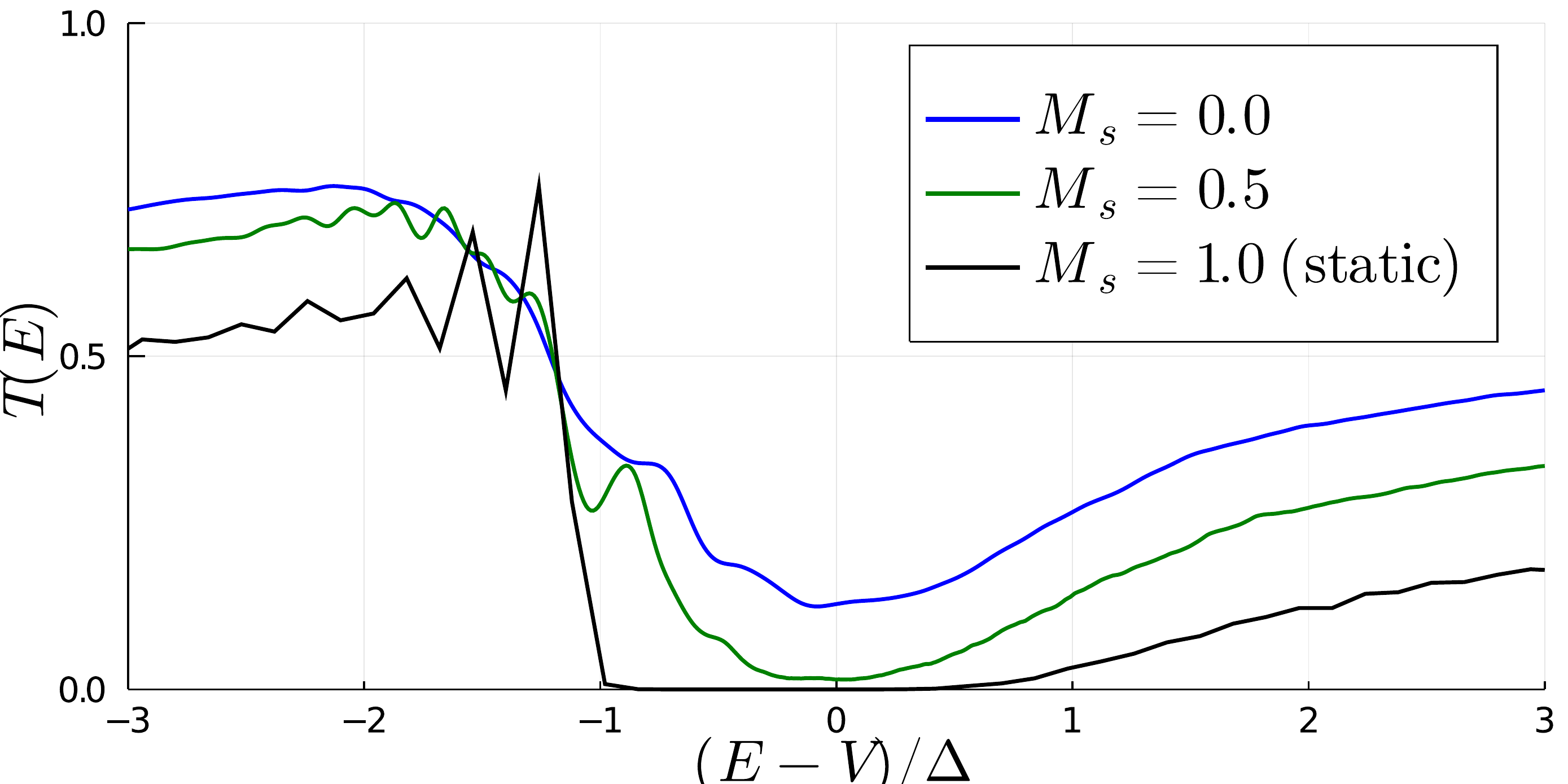}
	\caption{Transmission $T$ as a function of energy for $V/M=1.0$ and $\hbar\Omega/M=0.2$, $n_{max}=20$, $M_S=0.0$ (blue), and $M_S=0.5$ (green) compared to the static case $M_S=1.0$ (black). $|M|/\hbar\nu_F=0.2$ and $L/l_\Delta=8.0$ averaged over $N_{runs}=1000$ impurity configurations.}
	\label{fig:lfvsslow}
\end{center}
\end{figure}

In Fig.\ \ref{fig:lfvsslow} we compare the fully static case (black trace) to the slow driving case with $0\%$ (blue) and $50\%$ (green) static contribution.
We can see that the hard gap in the static case is not preserved in the dynamic case and the gap edge gets washed out.
The asymmetry in the static case is stronger than in the driven case.
Additionally, in this plot we can clearly see the reduced transmission in the static case.
This can be explained by the magnetic moment being approximately rotated into the $z$ direction for a short time such that in the adiabatic case no backscattering can take place.
Thus, averaged over a whole driving period, the transmission is higher than in the static case.

Throughout Figs.\ \ref{fig:hf}--\ref{fig:lf} we observe that a greater static contribution leads to a reduced transmission away from the center.
This confirms that even in the adiabatically driven case transport assisted by absorption contributes, regardless of the potential part.

\begin{figure}
	\begin{center}
		\includegraphics[angle=0,width=0.45\textwidth]{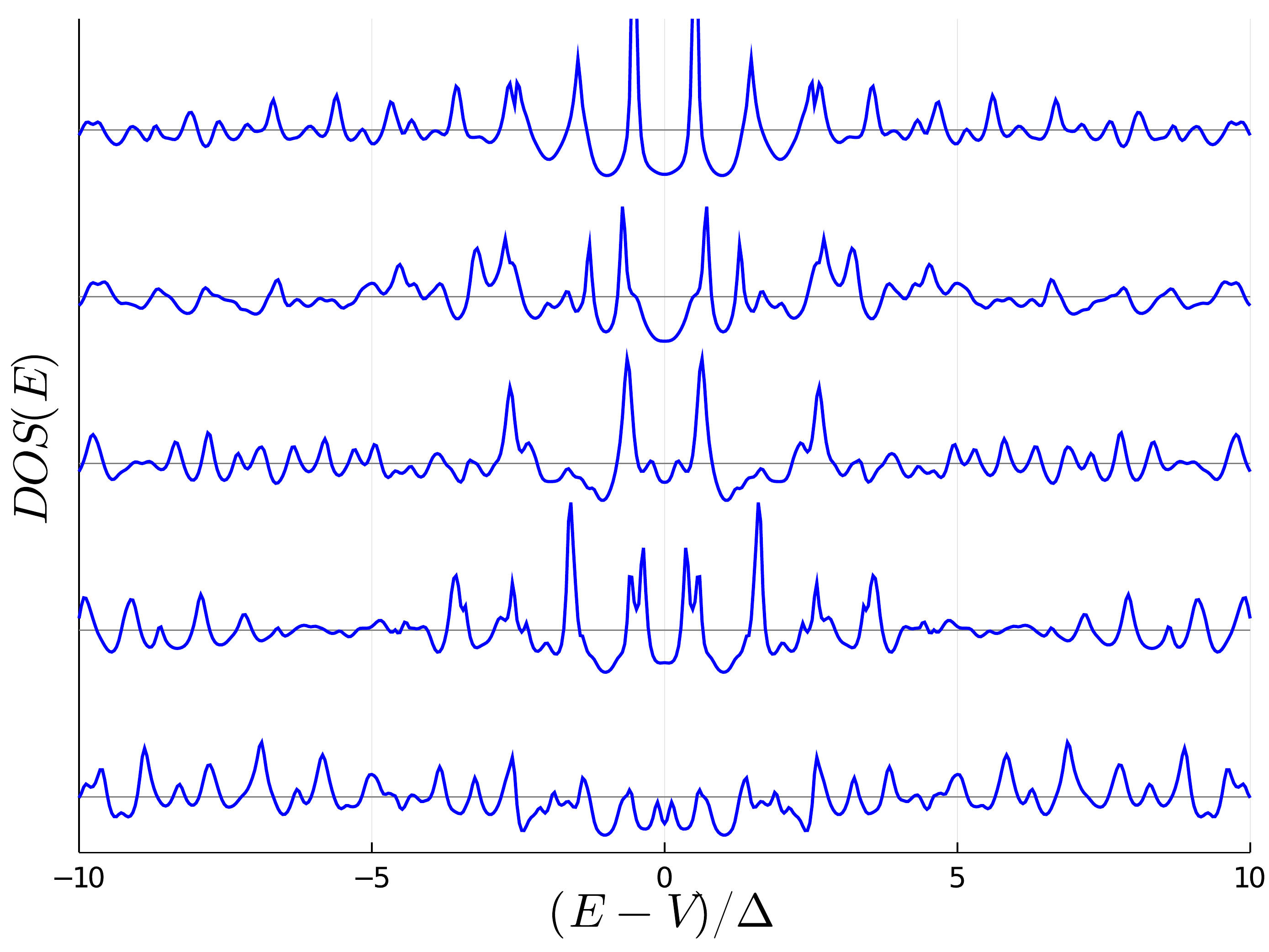}\\
		\includegraphics[angle=0,width=0.45\textwidth]{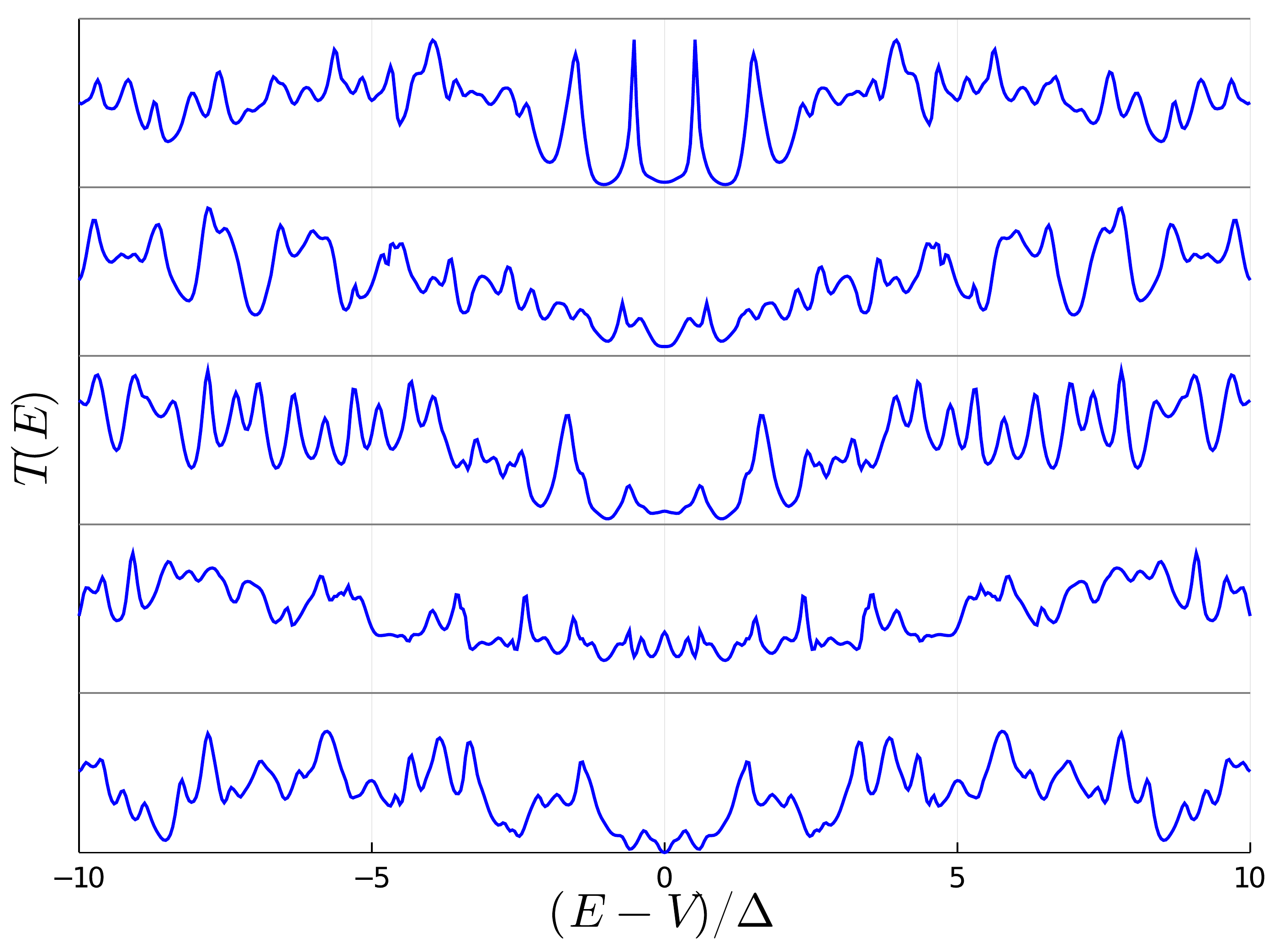}
		\caption{$DOS$ and transmission $T$ as a function of energy for single impurity configurations with $\hbar\Omega/M=2.0$, $n_{max}=6$, $M_S=0.0$, $|M|/\hbar\nu_F=0.2$, and $L/l_\Delta=8.0$.}
		\label{fig:t_single}
	\end{center}
\end{figure}

Up to now we have considered impurity averaged results.
In fig. \ref{fig:t_single} we present a set of traces from calculations for five separate impurity configurations at intermediate driving without any impurity averaging, where the same position of the trace in the DOS and the transmission correspond to the same impurity configuration.
As can be seen the features are mostly already visible for single impurity configurations, more so in the DOS.
Nevertheless, there are configurations that do not clearly show the resonance peaks visible in the impurity averaged results.

\section{Summary and conclusion}
	\label{sec:summary}

	We have developed a matrix GF for harmonically driven impurities.
	The formalism was applied to the case of magnetically aligned impurities with harmonically rotating magnetic moments placed on the edge of a 2D TI in a finite interval.
	Having calculated the GF we extract the DOS directly and the transmission via a scattering formalism that allows one to extract the full Floquet scattering matrix.
	The DOS and transmission have been calculated in three regimes of different driving frequency.
	We can explain the DOS profiles by considering the Floquet band structure of a system with a homogeneous rotating magnetic field as a reference.
	A common feature of all driving regimes is the flattening of the DOS with increasing potential part and approaching the clean system value.
	In all cases an asymmetry in the DOS and transmission introduced by the potential part can be observed, similar to the static impurity case, but less prominent.
	While the transmission also flattens it does not approach the clean system value, indicating that the potential part cannot screen the impurities in a way that fully prevents backscattering.
	Our results can be of value for future applications relying on engineering of edge state properties.
	
	\begin{acknowledgments}
	
	S.I.E. acknowledges support from the Reykjavik University Research Fund; M.L. and S.W. acknowledge support from the Knut och Alice Wallenbergs Stiftelse (Project No. 2016.0089) and from the Vetenskapsrådet (VR).
	\end{acknowledgments}

\appendix

\section{Fourier transform of the two-time Green's function}
\label{app:FT_GF}
In this appendix we present the derivation of Eq. \eqref{eq:scatsol} in the main paper.     We start from the scattering solution in time domain
    \begin{equation}
		\Psi_{\alpha,s}^{(+)}(x,t)=- \mathrm{i} \hbar \nu_F \left[ \int \mathrm{d}t'G(t,t';x,x') \sigma_z \Phi_{\alpha,s}^{(+)}(x',t') \right]_{x'=0}^{L}.
    \end{equation}
    and Fourier transform with respect to the time $t$.
    The only $t$-dependent part is the integral, so we will focus on that.
    First, we shift the integration variable $t'\rightarrow t-t'$, $\mathrm{d}t'\rightarrow -\mathrm{d}t'$
    \begin{eqnarray}
		\int \mathrm{d}t'G(t,t';x,x') \sigma_z \Phi_{\alpha,s}^{(+)}(x',t') \nonumber \\
		= -\int \mathrm{d}t'G(t,t-t';x,x') \sigma_z \Phi_{\alpha,s}^{(+)}(x',t-t').\label{eq:shifted}
    \end{eqnarray}
    We can now use that $G(t,t-t';x,x')$ is periodic in $t$ with period equal to the driving period $\Omega$ and write it as a Fourier sum
    \begin{equation}
        G(t,t-t';x,x') = \sum_n e^{i n \Omega t} G_n(t';x,x').
    \end{equation}
    Inserting this in Eq.\ \eqref{eq:shifted} and rewriting $1=e^{i n \Omega t'}e^{-i n \Omega t'}$ under the integral leads to
    \begin{eqnarray}
      & & -\int \mathrm{d}t'G(t,t-t';x,x') \sigma_z \Phi_{\alpha,s}^{(+)}(x',t-t') \nonumber \\
      &=& -\sum_n \int \mathrm{d}t'e^{i n \Omega t'}G_n(t';x,x') \sigma_z e^{i n \Omega (t-t')}\Phi_{\alpha,s}^{(+)}(x',t-t') \nonumber \\
    &=& - \sum_n \left((e^{i n \Omega \bullet} G_n(\bullet;x,x')) * (\sigma_z e^{i n \Omega \bullet} \Phi_{\alpha,s}^{(+)}(x',\bullet))\right)(t),\nonumber \\ 
    \end{eqnarray}
    where we have written the integral as a convolution in the last line.
    We can now use the convolutional theorem $\mathcal{FT}(g*h)=\mathcal{FT}(g)\mathcal{FT}(h)$ and the energy shift property $\mathcal{FT}(e^{iE't}g)(E)=\mathcal{FT}(g)(E-E')$ to find the full Fourier transform of the scattering solution
    \begin{equation}
    	\Psi_{\alpha,s}^{(+)}(x,E)= \mathrm{i} \hbar \nu_F \sum_n \left[ G_n(E_n;x,x') \sigma_z \Phi_{\alpha,s}^{(+)}(x',E_n)\right]_{x'=0}^{L}.
    \end{equation}

\begin{figure}[b]
	\centering
	\includegraphics[width=0.45\textwidth]{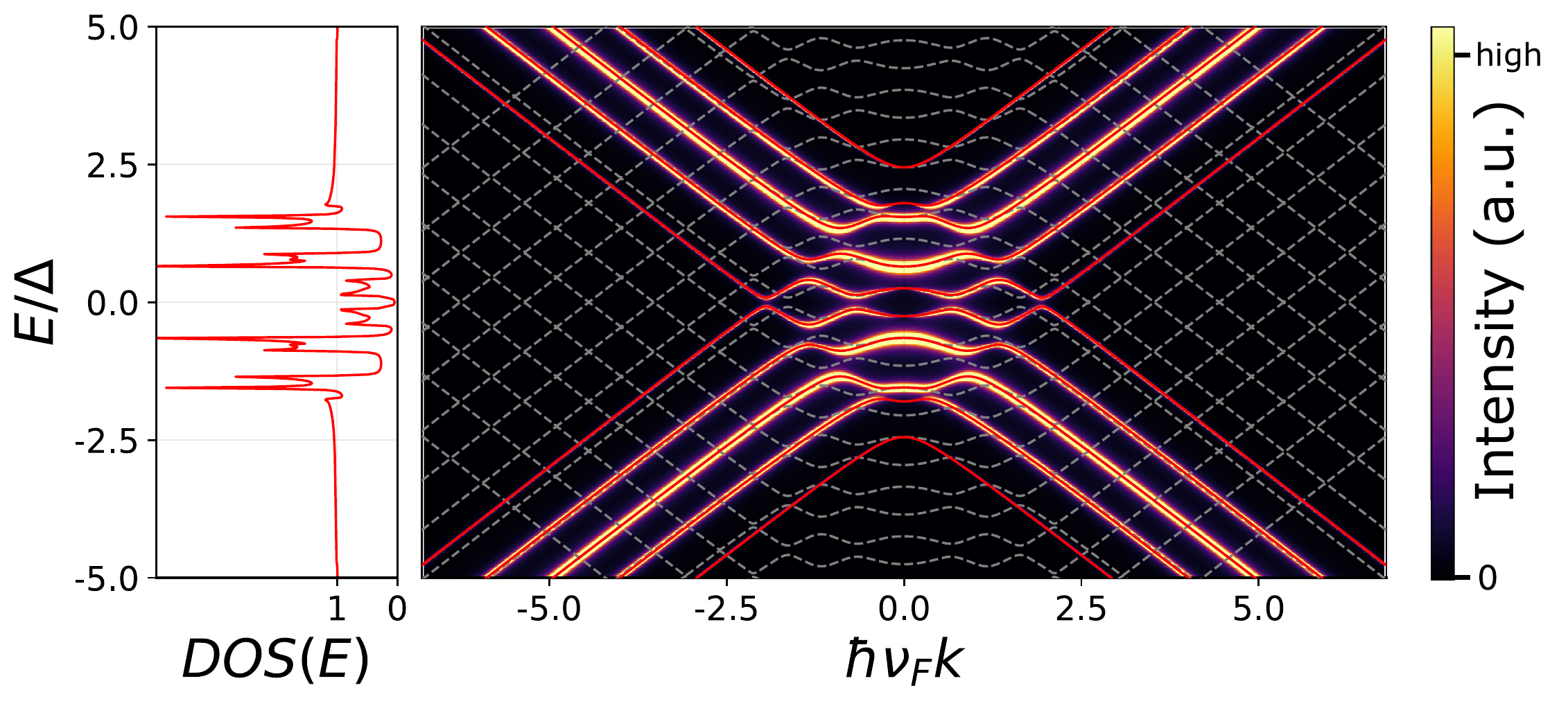}
	\caption{Right panel: band structure calculated directly from the Floquet-Hamiltonian (red and gray dashed lines) and directly by keeping the first two Fourier components (heat map) for a homogeneous dynamic magnetic barrier. Left panel: density of states via integration of the heat map. Floquet parameters: $\hbar\Omega/\Delta=0.9$, $n_{max}=2$ (red), and $n_{max}=10$ (gray).}
	\label{fig:fcestresbarrier09}
\end{figure}

\begin{widetext}
\section{Floquet-band structure for a homogeneous magnetic field}
\label{app:fbs}

The BHZ effective edge Hamiltonian with a homogeneous, magnetic field harmonically rotating in the $xz$ direction reads
\begin{equation}
H(t)=\hbar \nu_F k\sigma_z+ \Delta( \sin(\Omega t)\sigma_x + \cos (\Omega t)\sigma_z).
\end{equation}

In the basis of the helical edge states we can write the Floquet Hamiltonian in the matrix representation (for details, see Refs.\,\cite{thesissimon,Grifoni1998})

\begin{equation}
\left(\begin{array}{c|c|c|c|c}
\ddots &\ddots & & & \\ \hline
\ddots & \hbar \nu_F k\sigma_z -\hbar\Omega\sigma_0 & -\frac{i}{2}\Delta (\sigma_x + i \sigma_z) & & \\ \hline
& \frac{i}{2}\Delta (\sigma_x - i \sigma_z) & \hbar \nu_F k\sigma_z & -\frac{i}{2}\Delta (\sigma_x + i \sigma_z) & \\ \hline
& & \frac{i}{2}\Delta (\sigma_x - i \sigma_z) & \hbar \nu_F k\sigma_z +\hbar\Omega\sigma_0 &\ddots \\ \hline
& & &\ddots &\ddots
\end{array}\right).
\end{equation}
The diagonal blocks correspond to identical copies of the linear edge state dispersion shifted by $\pm\hbar\Omega$ and the off-diagonal blocks introduce a coupling between these branches.
For practical purposes the Fourier series will be truncated at a finite component with index $n_{max}$, resulting in a $((2n_{max}+1) 2)\times ((2n_{max}+1) 2)$ matrix for the Floquet eigenvalue equation, i.e. $(2n_{max}+1)^2$ $2\times 2$ blocks, that can then be treated numerically.
In Fig.\ \ref{fig:fcestresbarrier09} we show the quasienergies depending on the momentum $k$ and thereby find the {\it quasi-band structure}.

If the Fourier series is truncated at $n_{max}$, higher bands and the coupling to them is not considered.
In Fig.\ \ref{fig:fcestresbarrier09} two cases, $n_{max} = 2$ (red) and $n_{max} = 10$ (gray), are shown.
Truncation leads to a violation of the periodicity condition $E_n=E_0-n\hbar \Omega$.
By carrying more Fourier components we can preserve at least approximate periodicity in the energy range we are interested in.
That means in order to get reasonable results we make sure to carry enough Fourier components such that the quasi-energies in the desired energy range are approximately periodic and cover the whole energy range.
This approximate periodicity is the criterion for the choice of $n_{max}$ throughout the article.

Alternatively to calculating the band structure directly, we can use the matrix Eq.\,\eqref{eq:matrixGF} and truncate it to calculate $G_0(k,E)$ explicitly.
For $n_{max}=2$ this gives
\begin{eqnarray}
G_0(k,E)={}&\left( 1-\tilde{g}(k,E)\left[ V_+ \frac{\tilde{g}(k,E+\hbar \Omega)}{1-\tilde{g}(k,E+\hbar \Omega)V_+\tilde{g}(k,E+2\hbar \Omega)V_+^\dagger} V_+^\dagger + V_+^\dagger \frac{\tilde{g}(k,E-\hbar \Omega)}{1-\tilde{g}(k,E-\hbar \Omega)V_+^\dagger\tilde{g}(k,E-2\hbar \Omega)V_+} V_+\right] \right)^{-1}\tilde{g}(k,E)
\end{eqnarray}
and  via the spectral function the band structure can be calculated.
This is shown as a heat-map-like background in  Fig.\ \ref{fig:fcestresbarrier09}.
Integration over the momentum $k$ then gives the density of states $DOS(E)$ seen on the left of Fig.\ \ref{fig:fcestresbarrier09}, which cannot be achieved as easily with the direct approach.
A more detailed discussion and plots for different parameters can be found in \cite{thesissimon,jupyternotebooks}.

\end{widetext}


\end{document}